\documentclass[letterpaper,10pt]{article} 

\usepackage{opticameet3} %
\usepackage{pgfplots}
\pgfplotsset{compat=1.18}
\usepgfplotslibrary{fillbetween,groupplots}
\usepackage{tikz}
\usetikzlibrary{arrows, arrows.meta}
\usetikzlibrary{shapes.geometric}
\usetikzlibrary{backgrounds}
\usepgfplotslibrary{colorbrewer}
\pgfplotsset{
    tick label style={font=\footnotesize}, %
    label style={font=\footnotesize},      %
    xlabel style={yshift=5pt},
    ylabel style={yshift=-5pt},
    notch plot/.style={thick},
    notch 0/.style={notch plot, Set1-A, mark=x},
    notch 3/.style={notch plot, Set1-B, mark=+},
    notch 6/.style={notch plot, Set1-C, mark=*},
    notch 10/.style={notch plot, Set1-D, mark=square*},
    notch 20/.style={notch plot, Set1-E, mark=diamond*, mark size=2.5},
    notch 30/.style={notch plot, Set1-G, mark=triangle*, mark size=2.5},
} 
\usepackage{multicol}

\newcommand\authormark[1]{\textsuperscript{#1}}

\usepackage{amsmath,amssymb}
\usepackage[colorlinks=true,bookmarks=false,citecolor=blue,urlcolor=blue]{hyperref} 

\begin{document}

\title{Gas Line Absorption Mitigation in Hollow-Core Fibre using Spectral Pre-Equalisation}

\author{Eric Sillekens,\authormark{*} and Ronit Sohanpal}

\address{Optical Networks Group, UCL (University College London), London, WC1E 7JE, UK}

\email{\authormark{*}e.sillekens@ucl.ac.uk} %

\begin{abstract}
We study the impact of CO\textsubscript{2} absorption on hollow-core fibre transmission. Using spectral pre-equalisation, we digitally post-compensate gas-line absorption and show a 5.5~dB reduction in Q-factor penalty, outperforming a 383-tap equaliser by 1.3~dB.
\end{abstract}

\section{Introduction}

Hollow-core fibres (HCFs) have been demonstrated to greatly outperform conventional silica single-mode fibres by offering ultra-low attenuation (\textless 0.05~dB/km), reduced latency ($\sim30$\%) and low chromatic dispersion ($<$5~ps/(nm$\cdot$km))\cite{petrovich2025first,goa202540km}. Consequently, there is considerable interest towards deploying HCF in long-haul transmission systems where the broadband low-loss window enables fewer repeaters than current deployed links \cite{Poggiolini2025}.

Recently, transmission experiments using HCFs have been demonstrated for both C-band and SCL-band transmission systems, using both offline DSP \cite{ospina2025ultrawideband} and real-time transceivers\cite{hong2025realtime}, showcasing the viability of HCF-enabled transmission across both short- and long-haul scenarios. However, a major limitation of HCFs is the gas-line absorption (GLA) arising from residual trapped atmospheric gases. These gases include carbon dioxide (CO\textsubscript{2}), carbon monoxide (CO) and water vapour (H\textsubscript{2}O), which introduce additional spectral absorption phenomena within the transmission band. Of these, CO\textsubscript{2} is the most dominant, causing sharp absorption peaks within the C-, L- and U-bands. CO\textsubscript{2}-induced GLA leads to localised impairments for any overlapping data channels, severely penalising SNR and channel throughput at long transmission distances \cite{ospina2025ultrawideband,chen2025characteristics,hong2025realtime}. The GLA lines within the channel bandwidth manifest as increased inter-symbol interference (ISI) resulting in a demonstrated SNR penalty of up to 6~dB\cite{ospina2025ultrawideband}. To maximise the notable advantages of HCFs, particularly the usable bandwidth around the 1550~nm region, it is necessary to develop methods to mitigate the penalties associated with GLA.  While CO\textsubscript{2} elimination has been previously demonstrated using fibre post-processing \cite{Xiong2025elimination}, this purging method is slow for fibre lengths greater than 5~km. DSP methods are most desirable due to their flexible application in scenarios where residual GLA impairments exists. Entropy-loaded OFDM has been shown to mitigate some of the GLA penalties versus single-carrier systems, but this requires accurate knowledge of both fibre GLA characteristics and the transmitter laser wavelengths \cite{Sampaio2025hollow}.

In this work, we investigate the impact of GLA in high baud rate hollow-core fibre transmission. Using transmission simulations, we show that in-band GLA absorption peaks apply a significant penalty to the channel Q-factor, requiring 383 equaliser taps to recover 4.2~dB Q-factor degradation. In addition, we propose a novel spectral pre-equalisation scheme to mitigate GLA, showing 1.3~dB improved Q-factor recovery versus adaptive equalisation only, while only requiring 3 equaliser taps for optimum performance. This work demonstrates the application and potential of DSP methods for addressing physical impairments in hollow-core fibre transmission systems.

\begin{figure}[b]
\begin{multicols}{2}
    \centering
    (a)\raisebox{-2.1cm}{%
\begin{tikzpicture}
\begin{axis}[
width=8cm,
height=3cm,
xmin=1520,xmax=1610,
grid=both,
ymin=0,ymax=30,
xlabel = {Wavelength (nm)},
ylabel = {$\propto\alpha$ (dB)},
]
    \addplot[Dark2-C, ycomb, mark=none] table[x index=0, y expr=30*\thisrowno{3}] {data/hitran_telecom_lines.txt};
\end{axis}
\end{tikzpicture} %
} 
    (c)\quad\quad\quad \raisebox{-0.8cm}{\includegraphics[width=6cm]{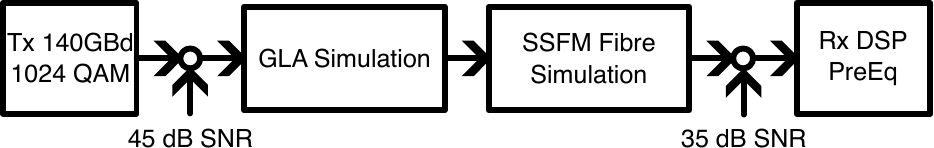}}
    (b)\raisebox{-3.4cm}{%
\begin{tikzpicture}
    \begin{axis}[
        width=7cm, height=4.5cm,
        xlabel={Frequency (GHz)},
        ylabel={$|H|^2$ (dB)},
        ymajorgrids,xmajorgrids,
        axis y line*=left,
        axis x line*=box,
        xmin=-5,xmax=10,
        legend pos=south west,
        legend style={nodes={scale=0.7, transform shape}},
        legend image post style={xscale=0.5},
    ]
    \addplot[Dark2-A, thick] 
        table[x index=0, y index=1] {data/lorentzian.txt};
         \addlegendentry{Amplitude}

    \draw[<->, thick] (axis cs:3,0) -- (axis cs:3,-10) node[midway,anchor=north east] {$A_i$};
    \draw[|-|, thick] (axis cs:4.5,-7) -- (axis cs:5.5,-7) node[pos=1,anchor=west] {$\Delta$};
    \end{axis}

    \begin{axis}[
        width=7cm, height=4.5cm,
        axis y line*=right,
        axis x line=none,
        ylabel={Phase (°)},
        ymajorgrids,
        xmin=-5,xmax=10,
        legend pos=south east,
        legend style={nodes={scale=0.7, transform shape}},
        legend image post style={xscale=0.5},
        ylabel style={yshift=10pt},
    ]
    \addplot[Dark2-B, thick]
        table[x index=0, y index=2] {data/lorentzian.txt};
        \addlegendentry{Phase}
    \end{axis}
\end{tikzpicture} %
} 
\end{multicols}\vspace{-.5cm}%
    \caption{(a) CO\textsubscript{2} gas line absorption from the HiTRAN database, (b) Bode plot of a simulated Lorentzian absorption peak, (c) HCF transmission simulation scheme.}
    \label{fig:lorentzian}
\end{figure}
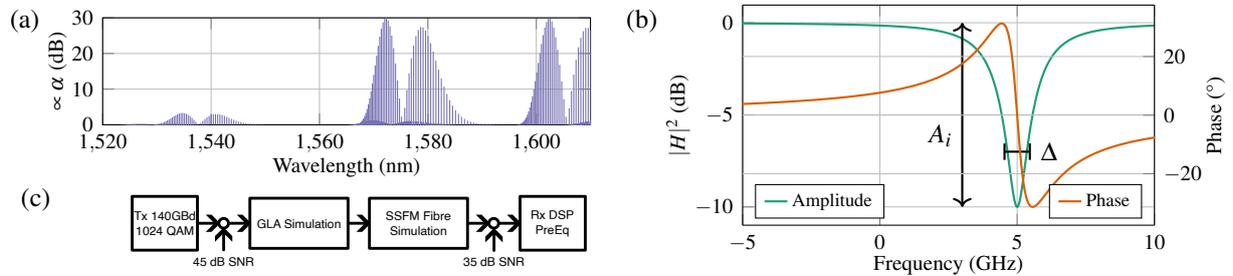

\section{HCF transmission simulation including GLA modelling}

We simulated a 300~km HCF link with CO\textsubscript{2} absorption. The HiTRAN CO\textsubscript{2} absorption spectrum is shown in Fig. \ref{fig:lorentzian}(a). GLA was modelled as series of Lorentzian-shaped notches in the frequency domain
\begin{equation}
    |H_{\mathrm{GLA}}(f)|=\prod_i\!\left(1-\frac{A_i}{1+\left(2/\Delta(f-f_i)\right)^2}\right),
\end{equation}
where $H_{\mathrm{GLA}}(f)$ is the GLA transfer function, $f_i$ is the centre frequency of notch $i$, $A_i$ is the $i$-th notch depth and $\Delta$ is the notch FWHM (assumed constant for all notches), shown in Fig. \ref{fig:lorentzian}(b). The amplitude and phase response of a single simulated notch is shown in Fig.~\ref{fig:lorentzian}. Since GLA is a causal phenomenon, we enforced causality via a minimum-phase complex response using the Hilbert transform to reconstruct the phase:
\begin{equation}
\phi(f) = -\mathcal{H}\left\{\ln|H_{\mathrm{GLA}}(f)|\right\},\qquad
H_{\mathrm{GLA}}(f)=|H_{\mathrm{GLA}}(f)|\,e^{j\phi(f)}. \label{eq:2}
\end{equation}

Notches were spaced every 40~GHz to emulate GLA line spacing with a FWHM of 1~GHz \cite{chen2025characteristics}. To simulate the worst-case penalty, we added a 5~GHz offset to the notch centre frequency to avoid a symmetric received spectrum; this ensures a complex GLA impulse response. The notch depths are varied to emulate different effective HCF lengths. For the transmission simulation, we generated a 140~GBd dual-polarisation 1024-QAM data channel with a 0.01\% root-raised-cosine filter and 100~kHz emulated linewidth, to which the GLA notch response was applied. In this case, 3 notches were present within the channel bandwidth. The split-step Fourier method was used to simulate HCF transmission using 5 steps, with a dispersion of 3~ps/(nm$\cdot$km), a loss of 0.05~dB/km and a nonlinear coefficient $\gamma =$1.2~mW/km. We neglect PMD here to focus on the performance impact of GLA only. Additional transmitter and receiver noise was modelled as AWGN with a transmitted/received OSNR of 45 and 35~dB per 12.5~GHz respectively. Finally, pilot-based DSP was used with a $2^{10}$-symbol pilot sequence to estimate the recursive least squares (RLS) and a pilot rate of $1/32$ \cite{Wakayama2021}.

\section{Spectral pre-equalisation scheme.}
In this work, we propose a frequency domain pre-equalisation technique to mitigate the GLA-induced penalty. We divided the data into 4096 sample blocks and estimated a two-sided PSD using Welch’s method with 50\% block overlap across the whole received sequence. We obtained the amplitude target of the pre-equaliser by taking the inverse square-root of the two-sided PSD $1/\sqrt{P_{\mathrm{Welch}}(f)}$.
To reconstruct the phase target of the pre-equaliser, we define the inverse square-root log-amplitude \( L(f) = -\tfrac{1}{2}\,\ln\!\big(P_{\mathrm{welch}}(f)\big) \) to compute the negated Hilbert transform. Comparing with Eq.~\eqref{eq:2}, the net phase response of the GLA after pre-equalisation will therefore be unity. We obtain our pre-equaliser frequency response as
\begin{equation}
H_{\mathrm{preEQ}}(f)=\exp\!\big(L(f) + j\mathcal{H}\{L(f)\}\big).
\end{equation}
where the first and second terms are the linear amplitude and phase response respectively. Out-of-band spectral bins were set to unity; the in-band mean magnitude was normalised to one. Finally, we apply a regulariser\cite{Salz1973Optimum} to the pre-equaliser to balance the impact of ISI and noise:
\begin{equation}
W_{\mathrm{mmse}}(f)=\frac{|H_{\mathrm{preEQ}}(f)|^2}{|H_{\mathrm{preEQ}}(f)|^2+\lambda}\in(0,1],\qquad
\tilde{H}_{\mathrm{preEQ}}(f) = H_{\mathrm{preEQ}}(f)\,W_{\mathrm{mmse}}(f).
\end{equation}
where $\lambda=10^{-\mathrm{SNR_{est}}/10}$, SNR\textsubscript{est} is the estimated channel SNR and $W_{\mathrm{mmse}}(f)$ is the regularisation factor. We used a fixed SNR estimate of 20~dB for all results presented in this work. A short complex adaptive equaliser (RLS at 2~samples/symbol) follows the spectral pre-equaliser. We sweep the tap count $N$ and report Q-factor calculated from the bit-error rate averaged over the whole transmission frame.

\section{Results and discussion}

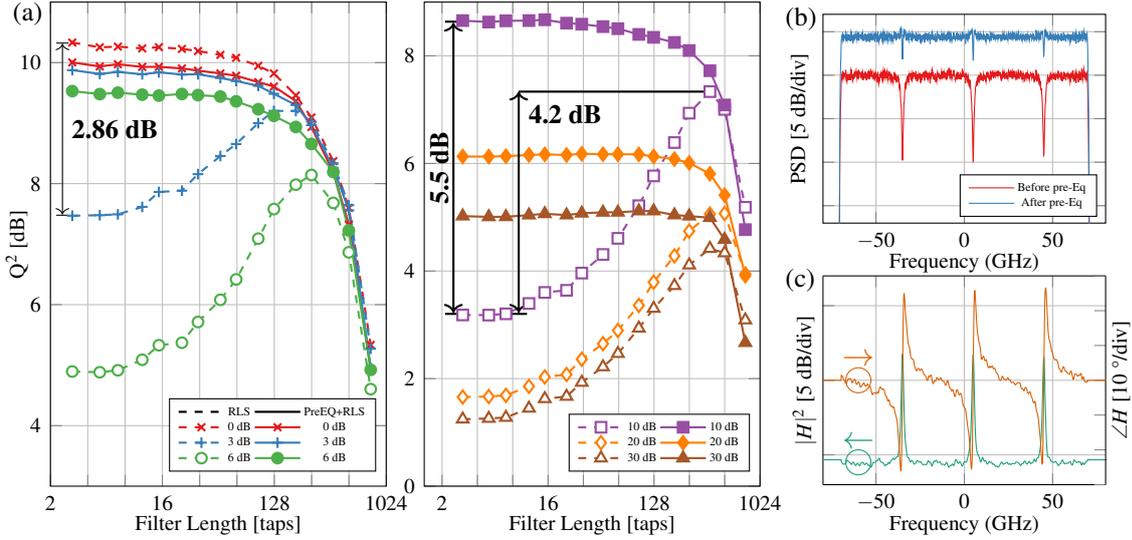
\begin{figure}[t]
    \centering
\begin{tikzpicture}
\begin{groupplot}[
group style={group size=2 by 1, horizontal sep=0.5cm},
  width=6cm,
  height=8cm,
  ]

\nextgroupplot[
  xlabel={Filter Length [taps]},
  ylabel={$\mathrm{Q}^2$ [dB]},
  legend pos=south east,
  legend columns=4,
  transpose legend,
  xmode=log,
  log basis x=2,
  legend style={font=\small, nodes={scale=0.5, transform shape}, xshift=0cm},
  ylabel style={yshift=-5pt},
  grid=both,
  xmin=2,xmax=1024,
  ymin=3,ymax=11,
  ytick distance=2,
  xtick={2,4,8,16,32,64,128,256,512,1024},  %
  xticklabels={2,,,16,,,128,,,1024},
]
\addlegendimage{dashed, notch plot}
\addlegendentry{RLS}
\pgfplotsinvokeforeach{0,3,6} {
   \addplot[notch #1, dashed, mark options={solid, fill=white}] table[x=FilterLength, y=#1dB] {data/140Gb_nonWhitening_Q.txt};
    \addlegendentry{#1~dB}
}

\addlegendimage{notch plot}
\addlegendentry{PreEQ+RLS}
\pgfplotsinvokeforeach{0,3,6} {
   \addplot[notch #1, mark options={solid}] table[x=FilterLength, y=#1dB] {data/140Gb_Whitening_Q.txt};
    \addlegendentry{#1~dB}
}

\draw[|<->|] (axis cs:2.5,7.4692) -- (axis cs:2.5,10.3315) node[midway,anchor=west,font=\bfseries] {2.86 dB}; 
  
\nextgroupplot[
  xlabel={Filter Length [taps]},
  legend pos=south east,
  legend columns=3,
  transpose legend,
  xmode=log,
  log basis x=2,
  legend style={font=\small, nodes={scale=0.5, transform shape}, xshift=0cm},
  grid=both,
  xmin=1.41,xmax=1024,
  ymin=0,ymax=9,
  ytick distance=2,
  xtick={2,4,8,16,32,64,128,256,512,1024},  %
  xticklabels={2,,,16,,,128,,,1024},
]

\pgfplotsinvokeforeach{10,20,30} {
   \addplot[notch #1, dashed, mark options={solid, fill=white}] table[x=FilterLength, y=#1dB] {data/140Gb_nonWhitening_Q.txt};
    \addlegendentry{#1~dB}
}

\pgfplotsinvokeforeach{10,20,30} {
   \addplot[notch #1, mark options={solid}] table[x=FilterLength, y=#1dB] {data/140Gb_Whitening_Q.txt};
    \addlegendentry{#1~dB}
}

\draw[|<->|, thick] (axis cs:2.5, 3.1837) -- (axis cs:2.5, 8.6555) node[midway,anchor=south,font=\bfseries,rotate=90, yshift=-2pt] {5.5 dB};
\draw[thick] (axis cs:381, 7.3354) -- (axis cs:9, 7.3354);
\draw[|<->, thick] (axis cs:9, 3.1837) -- (axis cs:9, 7.3354) node[anchor=west,font=\bfseries,pos=0.9] {4.2 dB};

\end{groupplot}
\node[anchor=north west, inner sep=0,xshift=-0.5cm, yshift=0cm] at (group c1r1.north west) {(a)};
\end{tikzpicture} %
\hspace{-0.2cm}%
\begin{tikzpicture}
    \begin{groupplot}[
    group style={
        group size=2 by 2,
        vertical sep=0.7cm,
    },
    width=5.3cm,
    height=4.35cm,
    ]
        \nextgroupplot[
        xmin=-80,xmax=80,
        ymin=-22,ymax=2,
        grid=both,
        xlabel={Frequency (GHz)},
        ylabel={PSD [5 dB/div]},
        yticklabels=\empty,
        ytick distance=5,
        legend style={font=\small, nodes={scale=0.5, transform shape}, xshift=0cm},
        legend pos=south east,
        ]

        \legend{Before pre-Eq, After pre-Eq}
        \addplot[Set1-A] table[x index=0, y expr=\thisrowno{1}-5] {data/spectrum.txt};
        \addplot[Set1-B] table[x index=0, y expr=\thisrowno{2}] {data/spectrum.txt};

        \nextgroupplot[group/empty plot]

        \nextgroupplot[ 
        name=bottomleft,
        xmin=-80,xmax=80,
        ymin=-2,ymax=12,
        axis y line*=left,
        axis x line*=box,
        grid=both,
        xlabel={Frequency (GHz)},
        ylabel={$|H|^2$ [5 dB/div]},
        yticklabels=\empty,
        ytick distance=5,
        legend style={font=\small, nodes={scale=0.5, transform shape}, xshift=0cm},
        legend pos=south east,
        ylabel style={yshift=-3pt},
        ]

        \addplot[Dark2-A] table[x index=0, y index=1] {data/H_preeq.txt};
        \node[Dark2-A, circle, draw, label={[Dark2-A,label distance=-3pt,font=\large]above:$\leftarrow$}] at (axis cs:-60,-0.5) {};
        \nextgroupplot[
        at={(bottomleft.south west)},
        anchor=south west,
        axis y line*=right,
        axis x line=none,
        xmin=-80,xmax=80,
        ymin=-30,ymax=30,
        grid=none,
        xlabel={Frequency (GHz)},
        ylabel={$\angle H$ [10 °/div]},
        yticklabels=\empty,
        ytick distance=10,
        legend style={font=\small, nodes={scale=0.5, transform shape}, xshift=0cm},
        legend pos=south east,
        ylabel style={yshift=13pt},
        ]

        \addplot[Dark2-B] table[x index=0, y index=2] {data/H_preeq.txt};
        \node[Dark2-B, circle, draw, label={[Dark2-B,label distance=-3pt,font=\large]above:$\rightarrow$}] at (axis cs:-60,-0) {};
    \end{groupplot}
\node[anchor=north west, inner sep=0, xshift=-0.5cm, yshift=0.1cm] at (group c1r1.north west) {(b)};
\node[anchor=north west, inner sep=0, xshift=-0.5cm, yshift=0.1cm] at (group c2r2.north west) {(c)};
\end{tikzpicture} %
    \vspace{-2mm}%
    \caption{(a) Q factor vs.\ tap length with/without the spectral pre-equaliser for 140~GBd 1024~QAM transmission system, (b) the PSD of the received signal before and after spectral pre-equalisation, and (c) the Bode plot $H_\text{preEQ}(f)$.}
    \label{fig:snr_taps}
\end{figure}

Fig.~\ref{fig:snr_taps}(a) shows Q-factor (dB) versus RLS filter length for different GLA notch depths with and without spectral pre-equalisation. Without any GLA notches (0~dB notch depth), increasing the RLS equaliser filter length leads to a degradation in Q-factor due to the increasing equaliser approximation error. Note there is no PMD implemented in this work, thus the optimum filter length is 3 taps. When the notch depth is increased, we can see that the performance with fewer taps is severely impacted. A notch depth of only 3~dB is sufficient to cause Q-factor degradation of 2.86~dB, caused primarily by ISI. This penalty can be reduced to 1.13~dB by increasing the filter length to 127 taps, however this comes at the cost of greatly increased computational complexity. Increasing the notch depth above 3~dB leads to several decibels of Q-factor penalty that is partially compensatable by increasing the filter length, but increasing adaptive equaliser approximation error limits the maximum recoverable Q-factor.

Fig.~\ref{fig:snr_taps}(b) shows the channel spectrum before and after spectral pre-equalisation is employed, while Fig.~\ref{fig:snr_taps}(c) shows the amplitude and response response of the pre-equaliser. At 0~dB notch depth the inaccuracy of the PSD estimation introduces a 0.3~dB penalty in Q-factor for all filter lengths. Nevertheless, for increasing notch depth the spectral pre-equaliser shows significant mitigation of the GLA-induced Q-factor penalty. For 10~dB notch depth, the spectral pre-equalisation scheme outperforms RLS-equalisation-only by 5.5~dB at 3 filter taps, and by 1.3~dB when the RLS-equalisation-only scheme is operated at optimum filter length (383 taps). This demonstrates the efficacy of the proposed scheme as well as the large reduction in computational complexity versus conventional equalisation. It should be noted that the scheme cannot fully mitigate the GLA response, both due to additional noise added by the spectral pre-equaliser notch estimation and the reduced OSNR within the notch bandwidths. When pre-equalisation is compared to the notch-less case without spectral pre-equalisation, there remains a penalty of 0.46~dB, 1.67~dB and 5.33~dB at optimum tap length for 3~dB, 10~dB and 30~dB notch depth respectively. Since the spectral pre-equaliser is in the frequency domain, this scheme could be integrated into the chromatic dispersion compensation DSP block to reduce implementation complexity. In this work, we applied the pre-equaliser to the entire received sequence, but in practice it is possible to use block-wise processing instead to emulate real-time DSP.

\section{Conclusion}

We investigate the impact of GLA on single-channel coherent transmission performance. When using recursive-least-squares adaptive equalisation, we show that 383 taps are required to reduce the GLA-induced Q-factor penalty by 4.2~dB for a GLA notch depth of 10~dB. Additionally, we propose a spectral pre-equalisation scheme that reduces the Q-factor penalty by a further 1.3~dB for a total 5.5~dB reduction, but requires only 3 filter taps, greatly reducing computational complexity versus RLS-based equalisation only.

\noindent\textbf{Acknowledgements:} This work was supported by EPSRC grant EP/R035342/1 TRANSNET. Eric Sillekens is supported by the Royal Academy of Engineering Research Fellowship.
\vspace{-0.2cm}
\section*{References}
\vspace{-0.2cm}
\begin{multicols}{2}
\bibliographystyle{opticajnl}
\footnotesize
\bibliography{refs}
\end{multicols}

\end{document}